\documentclass[preprint, superscriptaddress,  amssymb, amsfonts, floats, byrevetex, nofootinbib]{revtex4}
\usepackage{latexsym}
\usepackage{graphics}
\usepackage{graphicx}
\usepackage{epsfig}

\newcommand{\bmat}{\left(\begin{array}}
\newcommand{\emat}{\end{array}\right)}
\newcommand{\beq}{\begin{equation}}
\newcommand{\eeq}{\end{equation}}

\def\yzero{\smash{\hbox{$y\kern-4pt\raise1pt\hbox{${}^\circ$}$}}}
\def\p{\partial}
\def\a{\alpha}

\def\d{\delta}

\def\om{\omega}

\def\-{\hphantom{-}}

\def\s2{\frac{1}{\sqrt2}}

\def\beq{\begin{equation}}
\def\eeq{\end{equation}}
\def\beqa{\begin{eqnarray}}
\def\eeqa{\end{eqnarray}}

\def\IF{\relax{\rm I\kern-.18em F}}
\def\II{\relax{\rm I\kern-.18em I}}
\def\IP{\relax{\rm I\kern-.18em P}}

\def\Dsl{\,\raise.15ex\hbox{/}\mkern-13.5mu D} 

\def\IC{\bf C}
\def\IZ{\bf Z}

\def\z2z2{$\IC^3/(\IZ_2\times\IZ_2)$}

\def\a{\alpha}

\def\d{\delta}

\def\l{\lambda}

\def\n{\nu}

\def\p{\pi}

\def\s{\sigma}
\def\t{\tau}

\def\z{\zeta}

\def\G{\Gamma}

\def\bo{{\raise-.3ex\hbox{\large$\Box$}}}               
\def\face{{\raise.2ex\hbox{$\displaystyle \bigodot$}\mskip-2.2mu \llap {$\ddot
        \smile$}}}                                      

\def\leftrightarrowfill{$\mathsurround=0pt \mathord\leftarrow \mkern-6mu
        \cleaders\hbox{$\mkern-2mu \mathord- \mkern-2mu$}\hfill
        \mkern-6mu \mathord\rightarrow$}       
\def\dvec#1{\vbox{\ialign{##\crcr
        \leftrightarrowfill\crcr\noalign{\kern-1pt\nointerlineskip}
        $\hfil\displaystyle{#1}\hfil$\crcr}}}           

\def\beq{\begin{equation}}
\def\eeq{\end{equation}}

\def\beqx{\begin{displaymath}}
\def\eeqx{\end{displaymath}}

\def\beqa{\begin{eqnarray}}
\def\eeqa{\end{eqnarray}}
\def\NO{\nonumber}

\begin{document}
\title{
\normalsize \mbox{ }\hspace{\fill}
\begin{minipage}{12 cm}
{\tt UPR-1028-T, PUPT-2076, hep-th/0303083}{\hfill}
\end{minipage}\\[5ex]
{\large\bf
 Conformal Field Theory Couplings for
Intersecting D-branes on Orientifolds
\\[1ex]}}
\date{\today}
\author{Mirjam Cveti\v c}
\affiliation{School of Natural
Sciences, Institute for Advanced Studies, Princeton NJ 08540 USA\\
and\\
 Department of Physics and Astronomy, Rutgers University,
Piscataway, NJ 08855-0849 USA} \altaffiliation{On Sabbatic Leave
from the University of Pennsylvania}

\author{Ioannis Papadimitriou}
\affiliation{Princeton University, Princeton NJ 08540 USA}
\altaffiliation{Exchange Scholar from the University of
Pennsylvania}

\begin{abstract}We present  a conformal field theory
 calculation of four-point and three-point correlation
functions  for the bosonic  twist fields  arising at the
intersections of  D-branes    wrapping  (supersymmetric) homology
cycles of Type II orientifold compactifications. Both the quantum
contribution from local excitations at the intersections and the
world-sheet disk instanton corrections are computed.  As a
consequence we obtain the complete  expression for the Yukawa
couplings  of chiral fermions with the Higgs fields. The
four-point correlation functions in turn lead to the determination
of the four-point couplings in the effective theory, and may be of
phenomenological interest.

\end{abstract}\maketitle

\section{Introduction}
In the recent years, the intersecting  D-brane configurations have
played an important role in several areas. The most prominent one
is the construction of four-dimensional solutions
\cite{bgkl,afiru,bkl,imr,magnetised} of  Type II string theory,
compactified on orientifolds . In particular the appearance of the
chiral matter \cite{Bachas95,bdl}  at the brane intersection
provides a promising starting point to construct  models  with
potential particle physics implications \cite{Bachas95,bdl}.

 The model-building with
intersecting branes
 was   developed in a series of papers. In particular, non-supersymmetric
\cite{bgkl,afiru,bkl,imr,magnetised} (and subsequently explored in
\cite{bonn,bklo,bailin,kokorelis})
 and more recently,
supersymmetric \  \cite{CSU1,CSU2,CSU3,blumrecent,CPS,Honecker,CP}
constructions with quasi-realistic features of the Standard-like
and grand-unified models have been given.  One has a tremendous
freedom in the constructions of non-supersymmetric models, since
the Ramond-Ramond tadpole cancellation conditions can be satisfied
for many  brane configurations leading to the Standard-Model gauge
group and three families of quarks and leptons. However, the fact
that the theory is non-supersymmetric introduces the
Neveu-Schwarz-Neveu-Schwarz uncancelled tadpoles as well as the
radiative corrections of the string scale. (For the constructions
with intersecting D6-branes the string scale is necessarily of the
order of the Plank scale. However examples \cite{CIMD5} with
intersecting D5-branes have been given, where the string scale can
be as low as the TeV scale.)

On the other hand supersymmetric intersecting D-brane
constructions are extremely constraining. Nevertheless such
supersymmetric constructions  with intersecting D6-branes, which
have the Standard-like \cite{CSU1,CSU2,blumrecent,CP} and grand
unified model spectra \cite{CSU2,CPS}, have recently been
constructed. In particular, these models have an additional
quasi-hidden gauge sector that is typically confining which may
have interesting implications for the supersymmetry breaking
\cite{CLW}. Note however, that these models typically suffer from
additional exotics \cite{CLS1}. {Both supersymmetric and
non-supersymmetric constructions have adjoint matter associated
with each brane configuration, since the toroidal cycles wrapped
by the branes are {\it not} rigid}. Interestingly, the embedding
of supersymmetric four-dimensional models with intersecting
D6-branes has a lift \cite{CSU2,CSU3} into M-theory that
corresponds to the compactification of M-theory on a singular
$G_2$  holonomy manifold \cite{AW,Witten,aW,CSU1,CSU2}.

While  phenomenology of both non-supersymmetric
\cite{imr,kokorelis,CIM1} and supersymmetric \cite{CLS1,CLS2,CLW}
models has been addressed, the actual string calculations of the
couplings in this theory have been limited. While the tree-level
gauge couplings are relatively easy to determine and their
features have been studied, see e.g. \cite{CLS1} and references
therein. A calculation of gauge coupling threshold corrections
\cite{Lust} is also of  interest, since it could be compared to
the strong coupling limit of M-theory compactified on the
corresponding $G_2$ holonomy space \cite{FW}.

An important set of tree level calculations involves the
open-sector states that appear at the brane intersections. These
states include the chiral matter. In the supersymmetric
constructions the appearance of the full  massless chiral
supermultiplet is ensured there.  The couplings of most interest
are the three linear superpotential couplings, such as the
coupling of quarks and leptons to the Higgs fields. On the other
hand the four-point couplings are also of interest, since they
indicate the appearance of potentially other higher order terms in
the effective Lagrangian.

The calculations of couplings of states (tachyons)  appearing at
the non-supersymmetric intersections of branes  also has
interesting implications in the study of tachyon potential and the
phenomenon of tachyon condensation.  In particular for specific
T-dual models of [p-(p+2)] bound state configurations the
four-point calculations have been addressed in
\cite{Gava,Justin,Antoniadis}.

The purpose of this paper is to perform explicit string
calculations of the four-point and three-point correlation
functions associated with the states appearing at the
intersections of branes that wrap  cycles of the internal tori.
The non-trivial part of the calculation involves the evaluation of
the correlation functions of four (three) bosonic twist fields,
which  signify the  fact that the  states at the intersection
arise from the sector with twisted boundary conditions on the
bosonic (and fermionic) string degrees of freedom.  (For
supersymmetric cycles the physical massless states at the
intersection correspond to the  chiral supermultiplets.)  We
employ the techniques of conformal field theory, which are related
to the study of bosonic twist fields of the closed string theory
on orbifolds \cite{Dixon}. Similar techniques were employed in the
study of  Type II string theory for bound-states of p-(p+2) brane
sectors \cite{Gava,Justin,Antoniadis}.

Specifically, we focus on intersecting D-branes wrapping
factorizable N-cycles of $T^{2N}=T^2\times T^2\cdots$. Thus, in
each $T^2$ the D-branes wrap one-cycles, and the problem reduces
to a calculation  of  correlation functions of bosonic twist
fields associated with the twisted sectors at intersections of
D-branes  on a general $T^2$. Thus the final answer is a product
of contributions from each correlation function on each $T^2$.

We provide a general result for:

\begin{equation} \langle \sigma_\nu (x_1)\sigma_{-\nu} (x_2)\sigma_\nu
(x_3)\sigma_{-\nu} (x_4)\rangle, \label{nunu}
\end{equation}
which corresponds to the bosonic twist field correlation function
of states appearing at the intersection of two pairwise parallel
branes with  intersection angle $\pi\, \nu$ (See Figure 1).

In the  case of the twist fields appearing at the same
intersection, our result is interpreted in terms of the volume of
the torus, the lengths of the one-cycles $L_1$ and $L_2$ that each
set of branes wrap and the intersection numbers $I$. \footnote{A
special case, when the branes wrap the canonical cycles of a torus
with the complex structure $\nu$,
  a T-dual interpretation of this  correlation
function is that of the bosonic twist fields for D0-D2 brane with
the magnetic flux $B=\cot(\pi\nu)$.} We also address the case when
the twist fields  are  associated with   different intersections
of the two branes.  In particular we address in detail the
summation over the instanton sectors for such general cases.

The next calculation that we set out to do is that of the
four-point correlation function:

\begin{equation}
\langle\sigma_\nu (x_1)\sigma_{-\n} (x_2)\sigma_{-\l}
(x_3)\sigma_{\l} (x_4)\rangle, \label{nulambda}
\end{equation}
which corresponds to the bosonic twist field correlation function
of states appearing at the intersection of  two  branes
intersecting at respective angles $\pi\nu$ and $\pi \lambda$  with
the third set of parallel  branes (See Figure 2). This correlation
function is specifically suited for  taking the limit of $x_2\to
x_3$  which factorizes to a three point function associated with
the intersection of three branes. This latter result is
particularly interesting since it provides a key element in the
calculation of the Yukawa coupling.

In  this set of calculations we determine {\it both the classical
part and the quantum part} of the amplitude and thus obtain the
exact answer for the calculation.  In particular, the calculation
of the quantum part depends only on the  angles (and is thus
insensitive to the  scales of the internal space).
 On the other
hand the classical part carries information on the actual
separation among the branes  and the overall volume of $T^2$ as
well.

In particular the full expression (both classical and quantum
part) for the Yukawa couplings  for branes wrapping factorizable
cycles of $T^6$ is written as\footnote{See Note Added at the end of the paper.}
\beq
Y=\sqrt{2}g_0 2\p\prod_{j=1}^3\left[\frac{16\p^2 B(\n_j,1-\n_j)}{
    B(\n_j,\l_j)B(\n_j,1-\n_j-\l_j)}\right]^\frac14\sum_m\exp-\frac{A_j(m)}{2\p\a'}\eeq where
$A_j(m)$ is the area of the triangle formed by the three
    intersecting branes on the j-th torus and  $g_0=e^{\Phi/2}$, with $\Phi$ corresponding to the Type IIA dilaton.
The coupling is between two fermion  fields and a scalar field, i.e. the
massles states appearing at the respective intersections,  whose
kinetic energies are taken to be canonically normalized.


While we were in a process of  completing this work
 the paper \cite{CIM} appeared where a comprehensive analysis of the
classical part of the string contribution (disk instantons) to the
Yukawa coupling in models with intersecting branes on Calabi-Yau
manifolds was given, and extensive  explicit calculations  of  the
classical string contributions for models of intersecting branes
on toroidal orientifolds were presented.

Our work has certain overlap with  that of \cite{CIM}. In
particular, our work focuses only on models with branes wrapping
factorizable N-cycles of $T^{2N}=T^2\times T^2\cdots$. We evaluate
the  classical action contribution by  explicitly solving for the
classical solutions of the bosonic string  with the boundary
conditions governed by the locations of the D-branes. For the
special case of the three point function we therefore  also derive
the result of \cite{CIM} that the classical string contribution to
the three-point coupling involves a summation over the
$\exp(-A/2\pi \alpha')$, where $A$-corresponds to the area of the
 triangles associated with the intersections of the branes
in each $T^2$.  On the other hand, we have also determined the
quantum part of the correlation functions, thus obtaining the full
expression for the couplings.

The paper is organized as follows. In Section II we determine the
correlation function (\ref{nunu}). In Section III we calculate
(\ref{nulambda}) and factorize it on a three point function   to
determine the corresponding Yukawa coupling. Conclusions, that
include comments on generalizations of these calculations as well
as physical implications are given in Section IV.

\begin{figure}
\begin{center}
\scalebox{0.7}{\rotatebox{-90}{\includegraphics{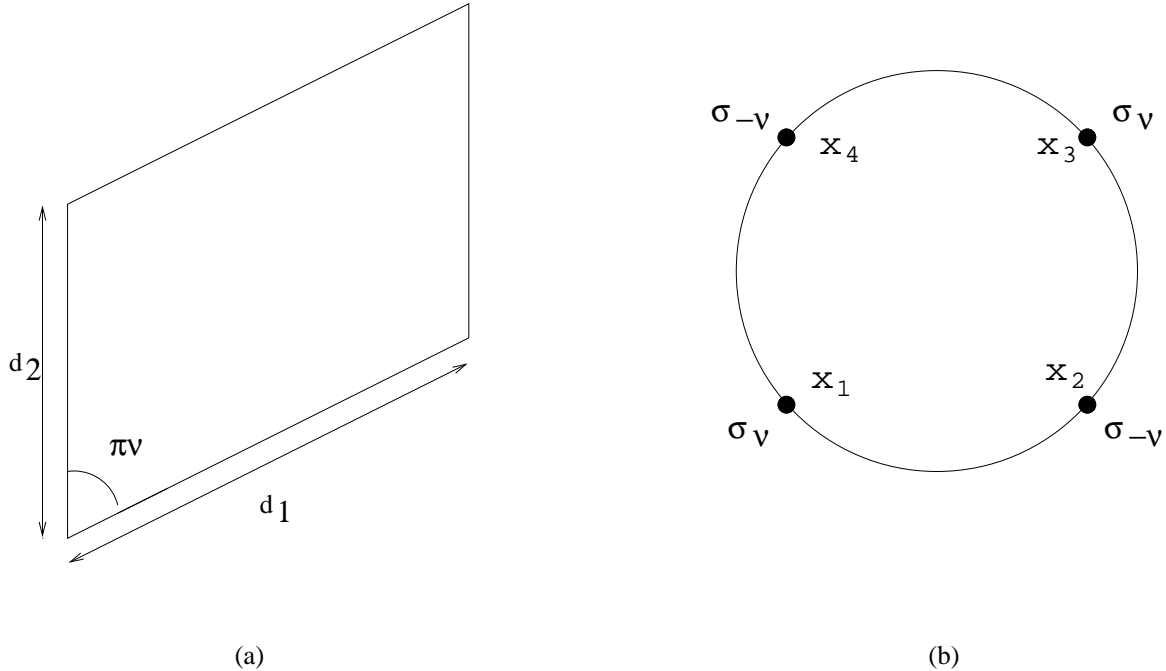}}}
\end{center}
\caption[]{\small  Target space: the  intersection of two parallel
branes separated by  respective distances $d_1$ and $d_2$ and
intersecting at angles $\pi \nu$ (Figure 1a).   World-sheet:
 a disk diagram of the
four twist fields located at $x_{1,2,3,4}$ (Figure 1b). The
calculation involves a map from the world-sheet to target space. }
\label{parallel}
\end{figure}

\begin{figure}
\begin{center}
\scalebox{0.7}{\rotatebox{-90}{\includegraphics{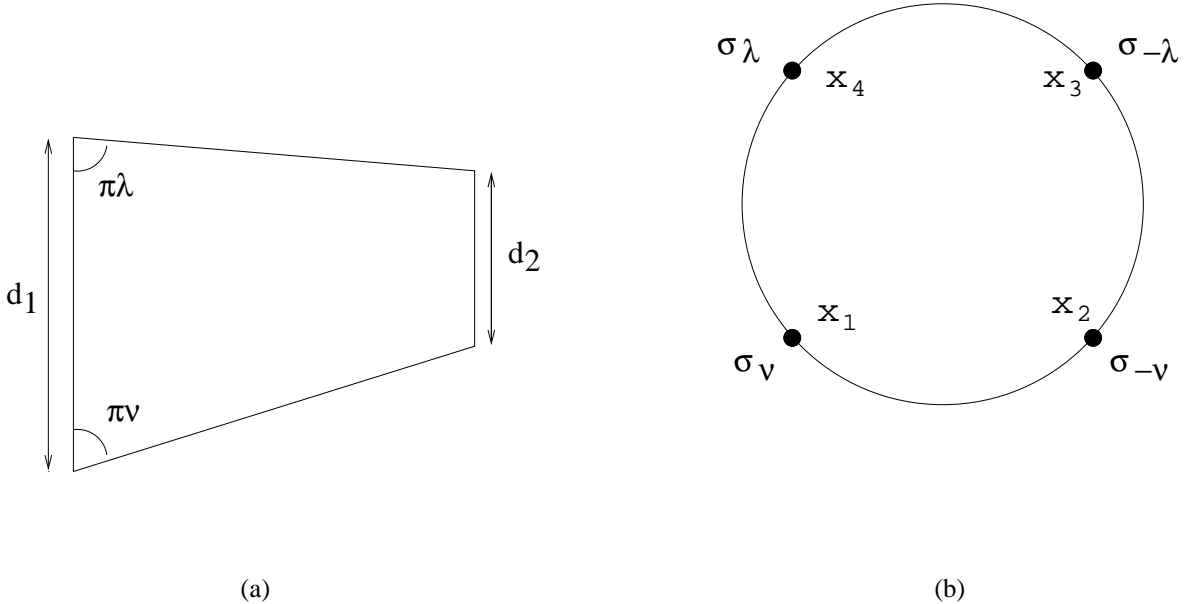}}}
\end{center}
\caption[]{\small  Target space: the  intersection of two branes
intersecting respectively with the two parallel branes at angles
$\pi \nu$ and $\pi \lambda$, respectively (Figure 2a).
World-sheet: a disk diagram of the four twist fields located at
$x_{1,2,3,4}$ (Figure 2b).The calculation involves a map from the
world-sheet to target space, allowing for a  factorization on
three-point function. } \label{nonparallel}
\end{figure}

\section{Four-Point Function with One Independent Angle}

In order to evaluate the path integral for the partition function
of open strings stretching between D-branes intersecting at an
angle $\p\n$ we split the embedding fields $X^i=X^i_{cl}+X^i_{qu}$
into a classical solution to the equation of motion, subject to
the appropriate boundary conditions, and a quantum fluctuation.
The mode expansion for the quantum fluctuation is not integer
moded due to the boundary conditions. The vacuum of the $X^i$ CFT
is then created by primary fields $\s_\n$ acting on the
$SL(2,\mathbb{R})$ invariant vacuum. The partition function
naturally factorizes into a classical contribution due to
worldsheet instanton sectors and a quantum amplitude due to
quantum fluctuations.  In contrast to the instanton contribution,
the quantum amplitude contains no topological information about
the worldsheet embedding in target space, but it is still
essential for the complete determination of Yukawa couplings in a
general model with intersecting branes.

\subsection{Evaluation of Quantum Amplitude}

  We shall employ
 the stress tensor method   \cite{Dixon, Gava, Justin} to evaluate the quantum amplitude of
four twist operators. For oriented theories the twist operators
live on the boundary of the disk and change the boundary
conditions as we move along the boundary. The boundary conditions
are specified by the D-brane configuration in target space (See
Fig. \ref{parallel}). We concentrate on a single $T^2$ and
D1-branes wrapping one-cycles. The amplitude for branes wrapping
factorizable three-cycles on $T^6$ is then the product of the
amplitudes for the three $T^2$ factors.

In terms of the complexified coordinates
$X=X^1+iX^2,\,\bar{X}=X^1-iX^2$ on $T^2$ the boundary conditions
read \beqa\label{bc}
\partial X+\bar{\partial}\bar{X}=0,\,\,
\partial\bar{X}+\bar{\partial}X=0,\,\,\,{\rm on}\,\,\,
(-\infty,x_1)\cup(x_2,x_3)\cup(x_4,+\infty) \\ \NO
e^{i\pi\nu}\partial X+e^{-i\pi\nu}\bar{\partial}\bar{X}=0,\,\,
e^{-i\pi\nu}\partial\bar{X}+e^{i\pi\nu}\bar{\partial}X=0,\,\,\,{\rm
on}\,\,\, (x_1,x_2)\cup(x_3,x_4). \eeqa These conditions define
the OPEs of the embedding fields with the twist operators, namely
\beqa \label{OPEs}\partial X(z)\s_\n(x)\sim(z-x)^{\n-1}\t_\n(x)+\ldots \\
\NO
\partial \bar{X}(z)\s_\n(x)\sim(z-x)^{-\n}\t'_\n(x)+\ldots \\ \NO
\bar{\partial}X(\bar{z})\s_\n(x)\sim-(\bar{z}-x)^{-\n}\t'_\n(x)+\ldots
\\ \NO
\bar{\partial}\bar{X}(\bar{z})\s_\n(x)\sim-(\bar{z}-x)^{\n-1}\t_\n(x)+\ldots
\eeqa and similarly for $\s_{-\n}(x)$. To evaluate the correlation
function of four twist fields
$\langle\s_\n(x_1)\s_{-\n}(x_2)\s_\n(x_3)\s_{-\n}(x_4)\rangle$ we
consider the correlators \beq g(z,w)=\frac{\langle
-\frac{1}{\a'}\partial
X(z)\partial\bar{X}(w)\s_\n(x_1)\s_{-\n}(x_2)\s_\n(x_3)\s_{-\n}(x_4)\rangle}
    {\langle\s_\n(x_1)\s_{-\n}(x_2)\s_\n(x_3)\s_{-\n}(x_4)\rangle}
    \eeq
\beq h(\bar{z},w)=\frac{\langle -\frac{1}{\a'}\bar{\partial}
X(\bar{z})\partial\bar{X}(w)\s_\n(x_1)\s_{-\n}(x_2)\s_\n(x_3)\s_{-\n}(x_4)\rangle}
    {\langle\s_\n(x_1)\s_{-\n}(x_2)\s_\n(x_3)\s_{-\n}(x_4)\rangle}
    \eeq
\beq \bar{g}(z,w)=\frac{\langle -\frac{1}{\a'}\partial
\bar{X}(z)\partial\bar{X}(w)\s_\n(x_1)\s_{-\n}(x_2)\s_\n(x_3)\s_{-\n}(x_4)\rangle}
    {\langle\s_\n(x_1)\s_{-\n}(x_2)\s_\n(x_3)\s_{-\n}(x_4)\rangle}
    \eeq and
\beq \bar{h}(\bar{z},w)=\frac{\langle -\frac{1}{\a'}\bar{\partial}
\bar{X}(\bar{z})\partial\bar{X}(w)\s_\n(x_1)\s_{-\n}(x_2)\s_\n(x_3)\s_{-\n}(x_4)\rangle}
    {\langle\s_\n(x_1)\s_{-\n}(x_2)\s_\n(x_3)\s_{-\n}(x_4)\rangle}.
    \eeq The OPEs (\ref{OPEs}) together with the conditions \beq
    g(z,w)\sim (z-w)^{-2},\,\,\,\,\, h(\bar{z},w)\sim {\rm
    regular}\eeq as $z\rightarrow w$ uniquely determine \beqa
    g(z,w)=\om_{1-\n}(z)\om_{\n}(w)\left[(1-\n)\frac{(z-x_1)(z-x_3)(w-x_2)(w-x_4)}{(z-w)^2}+\right.
    \\ \NO
    \left.\n\frac{(z-x_2)(z-x_4)(w-x_1)(w-x_3)}{(z-w)^2}+A(\{x_i\})\right]\eeqa
    and \beq h(\bar{z},w)=-\om_\n(\bar{z})\om_\n(w)B(\{x_i\})\eeq where
    \beq
    \om_\n(z)=(z-x_1)^{-\nu}(z-x_2)^{\nu-1}(z-x_3)^{-\nu}(z-x_4)^{\nu-1}.\eeq
    Here $A(\{x_i\})$ and $B(\{x_i\})$ are functions of the twist field
    positions to be determined. The boundary conditions and
    holomorphicity imply \beq
\label{holomorphicity}
    \bar{h}(z,w)=-g(z,w),\,\,\,\, \bar{g}(z,w)=-h(z,w).\eeq

    \begin{figure}
\begin{center}
\scalebox{0.7}{\rotatebox{-90}{\includegraphics{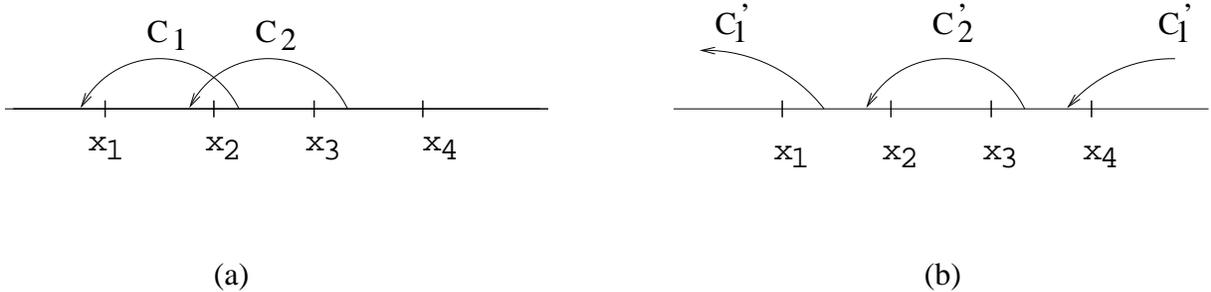}}}
\end{center}
\caption[]{\small World sheet contours. The contours $C_1$ and
$C_2$ (Figure 3a) are the two topologically inequivalent contours
leading to two independent conditions. The contours in Figure 3b
define the global monodromy conditions used in Section III. }
\label{contours}
\end{figure}

    In
    order to determine the functions $A$ and $B$ we
    impose appropriate monodromy conditions which will ensure that
    the quantum fluctuations $X_{qu}$ are local. This is
    guaranteed if \beq \int_{C_i}dX=\int_{C_i}d\bar{X}=0\eeq where
    $C_{i}$ is any non-trivial worldsheet contour. In the case at
    hand there are two topologically inequivalent contours $C_1$
    joining the intervals $(x_2,x_3)$ and $(-\infty,x_1)$ and
    $C_2$ joining the intervals $(x_3,x_4)$ and $(x_1,x_2)$ (See
    Fig. \ref{contours}a). One can save some effort, however, by noticing that the contours can be analytically
    continued to the worldsheet boundary along which only one
    particular linear combination of target space fields satisfies
    Neumann boundary conditions and can therefore have non-trivial
    displacement (along the boundary of the world sheet $d=d\t\partial_\t$  and
    hence the fields satisfying Dirichlet boundary conditions
    give no contribution). The non-trivial conditions are \beq
    \int_{C_1}d(e^{i\p\n}X-e^{-i\p\n}\bar{X})=0 \eeq \beq\label{cont2}
    \int_{C_2}d(X-\bar{X})=0. \eeq When inserted into the four twist field correlation functions  these lead to monodromy conditions for the
    correlators $g$, $\bar{g}$, $h$ and $\bar{h}$. For example, condition (\ref{cont2}) implies \beq
\int_{C_2}dz[g(z,w)-\bar{g}(z,w)]+\int_{C_2}d\bar{z}[h(\bar{z},w)-\bar{h}(\bar{z},w)]=0.\eeq Now, using relations (\ref{holomorphicity}) we can trade $h$ and $\bar{h}$ for
$g$ and $\bar{g}$. Moreover, $\bar{z}$ can be traded for $z$ by integrating along the mirror image of contour $C_2$ about the real axis, call it contour $\tilde{C}_2$. Taking into
 account the phases of $g$ and $\bar{g}$ on each of the contours one sees \beq \int_{\tilde{C}_2}dzg(z,w)= \int_{C_2}dzg(z,w),\eeq
 \beq \int_{\tilde{C}_2}dz\bar{g}(z,w)= \int_{C_2}dz\bar{g}(z,w).\eeq and hence  \beq \label{qmonodromy2}\int_{C_2}dz[g(z,w)-\bar{g}(z,w)]=0. \eeq
 Similarly we derive
\beq
    \label{qmonodromy1}
    \int_{C_1}dz[e^{i\p\n}g(z,w)-e^{-i\p\n}\bar{g}(z,w)]=0.\eeq

    Invoking
    $SL(2,\mathbb{R})$ invariance we fix $x_1=0,\,\,
    x_2=x,\,\, x_3=1,\,\, x_4\rightarrow\infty$. Dividing then by
    $\om_\n(w)$ and letting $w\to\infty$ we get \beq
    g(z,w)\rightarrow[\n(z-x)+A(x)]\tilde{\om}_{1-\n}(z) \eeq\beq
    \bar{g}(z,w)\rightarrow B(x)\tilde{\om}_{\n}(z)\eeq where
    $A(x)$ and $B(x)$ have been redefined appropriately to absorb
    the singularity arising from $x_4\to\infty$ and \beq
    \tilde{\om}_\n(z)=(z-x_1)^{-\nu}(z-x_2)^{\nu-1}(z-x_3)^{-\nu}.\eeq
    Conditions (\ref{qmonodromy1}) and (\ref{qmonodromy2})
    then give
 \beq \n\int_{C_2}dz(z-x)\tilde{\om}_{1-\n}(z)+A(x)\int_{C_2}dz\tilde{\om}_{1-\n}(z)-B(x)\int_{C_2}dz\tilde{\om}_{\n}(z)=0\eeq and
 \beq \n\int_{C_1}dz(z-x)\tilde{\om}_{1-\n}(z)+A(x)\int_{C_1}dz\tilde{\om}_{1-\n}(z)-e^{-2\p i\n}B(x)\int_{C_1}dz\tilde{\om}_{\n}(z)=0.\eeq Evaluating the contour integrals
 we find \beq -\n\int_{C_i}dz(z-x)\tilde{\om}_{1-\n}(z)=x(1-x)\frac{d}{dx}\int_{C_i}dz\tilde{\om}_{1-\n}(z),\eeq
\beq   \int_{C_1}dz\tilde{\om}_{1-\n}(z)=F(x)\eeq and \beq \int_{C_2}dz\tilde{\om}_{1-\n}(z)=e^{i\p\n}F(1-x)\eeq
    where \beq F(x)\equiv
   B(\n,1-\n)F(\n,1-\n;1;x)=\int_0^1dy
    y^{-\n}(1-y)^{\n-1}(1-xy)^{-\n}.\eeq  $B(\n,1-\n)$
    is the  Euler Beta function and  $F(a,b;c;x)$ is the Hypergeometric function.
 Solving for $A(x)$ we finally obtain
 \beq A(x)=\frac 12 x(1-x)\partial_x\log[F(x)F(1-x)].\eeq

    The quantum contribution $Z_{qu}(x)$ to the four twist field correlation function can now be
    extracted from the OPE\beq  \langle
    T(z)\rangle=\frac{h_\s}{(z-x)^2}+\frac{1}{z-x}\partial_x\log Z_{qu}(x)+\ldots
    \eeq as $z\to x$. Evaluating the left hand side by taking the
    limit $z\to x$ in the expression
    \beqa
    \langle T(z)\rangle=\lim_{w\rightarrow
    z}\left(g(z,w)-\frac{1}{(z-w)^2}\right)=\frac{A(\{x_i\})}{(z-x_1)(z-x_2)(z-x_3)(z-x_4)}\\
    \NO +\frac 12
    \n(1-\n)\left(\frac{1}{(z-x_1)}+\frac{1}{(z-x_2)}+\frac{1}{(z-x_3)}+\frac{1}{(z-x_4)}\right)^2.\eeqa
    we obtain
    \beqa
    Z_{qu}(x)=\lim_{x_4\rightarrow\infty}|x_4|^{\n(1-\n)}\langle\s_\n(0)\s_{-\n}(x)\s_{\n}(1)\s_{-\n}
    (x_4)\rangle=\\\NO {{\rm
    const.} \over {[x(1-x)]^{\n(1-\n)}[F(x)F(1-x)]^{1/2}}}.\phantom{more space}\label{Zqu} \eeqa

    \subsection{Evaluation of Classical Contribution}

    The path integral over the target space fields $X^i$ includes
    a sum over topologically inequivalent configurations from
    strings wrapping around the compact directions of the torus.
    The main contribution comes from configurations $X_{cl}^i$ satisfying the
    classical equations of motion while the effect of fluctuations
    about these classical configurations is encoded in $X_{qu}$
    and was calculated in the previous section using conformal
    field theory techniques.

    In this section we first determine the
    classical configurations satisfying the equation of motion subject to the boundary
    conditions (\ref{bc}), dictated by the D-brane setup. This is a straightforward boundary
    value problem for the Laplace operator in two dimensions and the
    solutions can be expressed in terms
    of holomorphic or antiholomorphic maps from the disk onto the
    target space manifold. We then evaluate the on-shell action
    and sum over the toroidal lattice to obtain the world-sheet instanton
    contribution to the four point function.

    The solutions to the above boundary value problem are \beqa
    \partial
    X(z)=a\om_{1-\n}(z)\equiv\tilde{a}e^{-i\p\n}\tilde{\om}_{1-\n}(z)\\\NO
     \bar{\partial}
    \bar{X}(\bar{z})=-a\om_{1-\n}(\bar{z})\equiv-\tilde{a}e^{i\p\n}\tilde{\om}_{1-\n}(\bar{z})\\\NO
     \partial
    \bar{X}(z)=b\om_{\n}(z)\equiv\tilde{b}e^{i\p(\n-1)}\tilde{\om}_{\n}(z)\\\NO
     \bar{\partial}
    X(\bar{z})=-b\om_{\n}(\bar{z})\equiv-\tilde{b}e^{-i\p(\n-1)}\tilde{\om}_{\n}(\bar{z})\eeqa
    where the coefficients $a$ and $b$ are the only free
    parameters to be determined. These parameters reflect the freedom in specifying the length of
    the two independent sides of the parallelogram. The classical contribution to the path integral is then
    \beq Z_{cl}=e^{-S_{cl}} \eeq where \beqa
    S_{cl}=\frac{1}{4\p\a'}\int_{\mathbb{C}_+}d^2z(\partial X\bar{\partial}\bar{X}+\partial\bar{X}\bar{\partial}X)\\
    \NO
    =-\frac{1}{2\p\a'}\sin(\p\n)F(x)F(1-x)(\tilde{a}^2+\tilde{b}^2)\eeqa
    where we have used \beq
    \int_{\mathbb{C}_+}d^2z|\tilde{\om}_\n(z)|^2=\int_{\mathbb{C}_+}d^2z|\tilde{\om}_{1-\n}(z)|^2=2\sin(\p\n)F(x)F(1-x).\eeq

    To determine the coefficients $\tilde{a}$ and $\tilde{b}$ we
    impose the monodromy conditions\footnote{We assume for simplicity that the branes wrap
    cycles along the two-torus lattice vectors.}  \beq
    \int_{C_1}ds=\frac{2n_1\p R_1}{\sin(\p\n)},\,\,\,\int_{C_2}ds=2n_2\p R_2 \eeq Since
    $X^2=\cot(\p\n)X^1$ along $C_1$ \beq
    ds^2=(dX^1)^2+(dX^2)^2=\left(\frac{dX^1}{\sin(\p\n)}\right)^2.\eeq
    Similarly, $ds^2=(dX^2)^2$ along $C_2$. A similar calculation
    as for the quantum monodromy conditions then gives \beqa
    \tilde{a}=i\p\left(\frac{n_1R_1}{\sin(\p\n)F(x)}+\frac{n_2R_2}{F(1-x)}\right)\\
    \NO
    \tilde{b}=i\p\left(\frac{n_1R_1}{\sin(\p\n)F(x)}-\frac{n_2R_2}{F(1-x)}\right).\eeqa
    Hence, \beq
    S_{cl}=\frac{2\p}{\a'}\sin(\p\n)F(x)F(1-x)\left[\left(\frac{n_1R_1}{\sin(\p\n)F(x)}\right)^2+
    \left(\frac{n_2R_2}{F(1-x)}\right)^2\right].\label{Scl}\eeq

The full amplitude  is now of the form:
\begin{equation}
Z(x)\equiv \lim_{x_4\to \infty}  |x_4|^{\nu(1-\n)} \langle
\sigma_\nu(0) \sigma_{1-\nu} (x) \sigma_\nu(1) \sigma_{1-\nu}
(x_4) \rangle =Z_{qu} \sum_{m_1,m2}e^{-S_{cl} (m_1,m_2)} \,
\label{Z}
\end{equation}
where  $Z_{qu}$ is determined in (\ref{Zqu}) (up to an overall
$const.$ and  $S_{cl}$   is defined in (\ref{Scl})).

 Note, in  the limit
    $R_1,\,R_2\to\infty$ \beq
    \sum_{m_1,m_2}e^{-S_{cl}(m_1,m_2)}\to 1\eeq and hence the
    four twist amplitude receives no instanton corrections as expected.

\subsection{Canonical Form of the Amplitude and Generalizations}
The above   calculation was performed for the
 amplitude of two intersecting branes wrapping two canonical homology cycles $[a]$ and $[b]$, respectively
 of  the $T^2$ with the complex structure specified by $\nu$. We can however reexpress this amplitude
 in  terms of a four-point twist amplitude for two branes wrapping  two general cycles  specified
 by the wrapping numbers $(n_1,m_1)$ and $(n_2,m_2)$ on  $T^2$ with
 the trivial complex structure, first, as:
 \beqa\label{Zo}
     Z(x)=const.[x(1-x)]^{-\n(1-\n)}[F(x)F(1-x)]^{-1/2}\\
     \NO
     \sum_{r_1,r_2}\exp-\frac{1}{2\p\a'}\sin(\p\n)F(x)F(1-x)\left[\left(\frac{r_1L_1}{F(x)}\right)^2+
    \left(\frac{r_2L_2}{F(1-x)}\right)^2\right].\eeqa
Where $L_i$ are the lengths of the corresponding one-cycles and
can be expressed in terms of the wrapping numbers and the radii of
the torus as: $L_i=2\p\sqrt{(n_iR_1)^2+(m_iR_2)^2}$.
  On the other hand $\sin(\p\n)$ can be reexpessed in terms of
    invariant quantities such as the intersection number $I_{12} \equiv  n_1m_2-n_2m_1$
    and the lengths $L_1,\,L_2$ of the
    one-cycles  as \beq
    \sin(\p\n)=\frac{(2\p)^2I_{12}R_1R_2}{L_1L_2}=\frac{I_{12}\chi}{\sqrt{n_1^2+\chi^2 m_1^2}\sqrt{n_2^2+\chi^2 m_2^2}}\eeq where $\chi\equiv R_2/R_1$ is the
    complex structure modulus.  As expected the angle is insensitive to the overall scale, and depends only
    on the wrapping numbers and the complex structure modulus.

It is  straightforward to generalize this result to a $T^2$ with a
non-trivial complex structure $\tau$. In this case it is efficient
to parameterize the  wrapping numbers in terms of $m_i\to {\tilde
m}_i\equiv m_i +\tau n_i$  (see for example \cite{CSU2}). The
complete result takes the form (\ref{Zo}), but with $m_i$'s
replaced with ${\tilde m}_i$.

Of course a generalization of the amplitude to the case of
$T^{2n}=T^2\times T^2\cdots $ (we assume the K\"ahler structure to
be a product of the K\"ahler structures of each $T^2$) is
straightforward.  In this case each twist field  is just a product
of individual twist fields for each $T^2$, and the four-twist
amplitude is a product of individual twist amplitudes (\ref{Zo}).
The most interesting examples where the above calculations can be
applied  are cases of Type IIA string theory on $T^6=T^2\times
T^2\times T^2$ with intersecting D6-branes wrapping a product of
three one-cycles associated with each $T^2$.

 For the purpose of performing complete string amplitude calculations it is instructive to
  write down the complete  vertex operators for physical  bosonic states  $\chi$ and $\chi^*$  which
  in
 the (-1) conformal ghost  ($\phi$) picture\footnote{Here for clarity we have denoted the bosonic antitwist
 operators by $\sigma_{1-\nu}$ modifying the notation $\sigma_{-\nu}$ which was used before for the same operators.}:
 \begin{equation}
 V_{-1;\chi}= e^{-\phi} \prod_{j=1}^3 \sigma_{1-\nu}^j
e^{i(1-\nu_j)H_j} e^{ik_\mu X^\mu}\, ,
  \ \ \
  V_{-1;\chi^*}= e^{-\phi} \prod_{j=1}^3 \sigma_\nu^j e^{-i(1-\nu_j)H_j}
  e^{ik_\mu X^\mu}\, ,\label{bos}
\end{equation}
  where  $H_i$ corresponds to the bosonized world-sheet fermion $\psi^i$ (worldsheet super-partner of the i-th toroidal
  coordinate $X^i$). Here we chose to write explicitly the  complete vertex
  operator for the bosonic states in four-dimensions; they  appear
  in the internal space
  at the intersection of D6-branes wrapping a product of
  three one-cycles  on $T^6=T^2\times T^2\times T^2$.
In the case of  supersymmetry  the  intersection angles  satisfy
the condition \cite{CSU2} $\sum_{j=1}^3\pi\nu_i=2\pi$ which
ensures that these vertex operators  correspond  to massless
bosonic states, which now become superpartners of  massless
fermionic states with the following (-1/2) superconformal ghost
picture vertex operators \cite{bdl}:
\begin{equation}
 V_{-\textstyle{1\over 2}; \chi}= e^{-\textstyle{\phi\over 2}}S_{\a} \prod_{j=1}^3 \sigma_{1-\nu}^j
e^{i(\textstyle{1\over 2} -\nu_j)H_j} e^{ik_\mu X^\mu}\, ,
  \ \ \
  V_{-\textstyle{1\over 2};\chi^*}= e^{-\textstyle{\phi\over 2}} \tilde{S}_{\a}\prod_{j=1}^3
   \sigma_\nu^j e^{-i(\textstyle{1\over 2}-\nu_j)H_j} e^{ik_\mu
   X^\mu}\, .
\label{fer}\end{equation} Here $S_{\a}=e^{\pm\textstyle{1\over
2}{\cal H}_1\pm \textstyle{1\over 2}{\cal H}_2}$ and
$\tilde{S}_{\a}=e^{\pm\textstyle{1\over 2}{\cal H}_1\mp
\textstyle{1\over 2}{\cal H}_2}$  represent the  spin fields with
respective positive and negative chirality. [$\sim e^{{\cal
H}_{1,2}}$ are bosonized worldsheet fermions $\psi^a$ with $a$
the four-dimensional (complexified) indices.]
  Note that in the case of supersymmetry the
vertex operators (\ref{bos}) for $\chi$ and $\chi^*$ have the
$N=2$ worldsheet charge $ H\equiv \sum_{i=1}^3 H_i$, $+1$ and
$-1$, respectively and thus correctly represent the vertex
operators for the bosonic component of the chiral superfield and
its complex conjugate, respectively
 \cite{DixonTriesteLectures}. Similarly, the
worldsheet charge $H$ for the fermionic vertex operators
(\ref{fer}) are  respectively $-{1\over 2}$ and ${1\over 2}$,
again in accordance with $N=2$ worldsheet supersymmetry
representing the fermionic components of the chiral superfield and
its complex conjugate, respectively.

In the above expressions we have suppressed the Chan-Paton
factors, however they are straightforward to incorporate. The
states transform as  $N\otimes\overline{M}$  under the $U(N)\times
U(M)$ gauge symmetry of the two intersecting branes (see, e.g.,
\cite{CSU2} for details).

 The  orientifold projection of Type IIA theory
 involves along with the  world-sheet parity projection also the  mirror symmetry projection, say along
 the horizontal $n$-plane of  $T^{2n}$. Note that  this projection restricts
 the value of $\tau$ to be either $0$ or $1\over
2$. Since each brane now also has an orientifold image, obtained
by a map $(n_i,{\tilde m}_i) \to (n_i,-{\tilde m}_i)$,  one can
now consider  four-amplitudes of   states appearing at the
intersection of a brane, denoted by i, and another one, denoted by
$j^*$, that is an orientifold image of a brane denoted by j. Note
that the calculation of the four-amplitudes proceeds analogously
as above. It is possible to calculate the four-point amplitude of
the states  appearing at the intersection of, say brane $i$, with
its own orientifold image $i^*$.   Such states can appear as
symmetric or anti-symmetric representations of the $U(N_i)$ (for
details see \cite{CSU2}). Again, calculation proceeds along the
same lines.


In the following we shall determine the crucial  normalization
$const.$ of the quantum part of the correlation function.

\subsection{Normalization of the  Amplitude}

    The overall normalization of the four-point amplitude can be
    determined  by factorizing the amplitude in the limits $x\to 0$ or $x\to 1$ in which the
    four-point amplitude reduces to  a product of the two three-point
    amplitudes. Namely in these limits, the four-point amplitude
     contains a dominant  contribution from the exchanges of the
    intermediate open string winding states around the compact
    directions.  In the effective field theory description the
    zero winding states correspond to the exchange
    of gauge bosons living on the brane along the cycle which is not collapsed by
    the limiting process. As $x\to 0$ brane 2 contributes while as
    $x\to 1$ brane 1 contributes.

For the sake of concreteness we shall focus on a four-dimensional
example with D6-branes wrapping a product of three-cycles. The
physical states at the intersections are represented by the vertex
operators  (\ref{bos}) and (\ref{fer}).
    For that purpose we shall evaluate the  four-point disk
    amplitude  $S_4(k_1,k_2,k_3,k_4)$   of  two   bosonic states
    $\chi$ and two  $\chi^*$
    at the intersections. We shall relate this
    amplitude to  the product of two three point functions  $S_3(k_1,k_2;k_3)$ of
    $\chi$  and   $\chi^*$  states, and the gauge boson $A_\mu$ via the unitarity condition:
  \begin{equation}
  S_4(k_1,k_2,k_3,k_4)=i [g_{YM}^0]^2\int {{d^4 k}\over {(2\pi)^4}}{{S_3(k_1,k_2;k)S_3(k_3,k_4;-k)}\over
  {-k^2+i\epsilon}}\, \label{unit}.
  \end{equation}

With these preliminaries  we now proceed with the calculations of
the physical string  amplitudes.   The three-point amplitude takes
the form:
\begin{equation}
 S_3 (k_1,k_2;k_3)\equiv \langle V_{-1;\chi} V_{-1;\chi^*}
V_{0,A_\mu}\rangle = i{{C_{D2}g_0^3}\over \sqrt{2\alpha'}}
(2\pi)^4\delta^{(4)}(\sum_{i=1}^3k_i) \alpha' (k_1-k_2)\cdot e \,
,
\end{equation}
This is the standard three point amplitude, since   we have taken
into account that $\langle \sigma_{\nu} (0) \sigma_{1-\nu}
(1)\rangle ={\bf 1}$, and that in the picture changing procedure
of the gauge-boson  vertex the internal part of the fermionic
stress energy tensor does not contribute.  We have introduced the
disk coupling $C_{D2}$ and the coupling $g_0$ of each vertex
operator. The additional factor of $(2\alpha')^{-1 /2}$ is due to
the picture changing procedure of the   gauge field vertex. The
gauge-field  polarization vector is denoted by $e_\mu$. (For
simplicity we calculated the amplitude only for $U(1)$ gauge
field;  generalization to $U(N)$ is straightforward.) The
factorization of the four-point gauge boson amplitude onto the
product of two three-point gauge boson amplitude yields the
standard relationship between $C_{D2}$ and $g_0$: $C_{D2}=1/(g_0^2
\alpha')$.   Note that this relationship also automatically
ensures that on both sides of (\ref{unit}) the dependence on $g_0$
drops out; namely $[C_{D2}g_0^3]^2=C_{D2}g_0^4/\a'$.

We have chosen $g_0=e^{\Phi/2}$ which allows one to write the
effective kinetic energy action for the gauge fields with the
pre-factor $1/[g_{YM}^0]^2$ and the  kinetic energy for the $\chi$
fields to be canonical. Here $g_{YM}^0$  is defined in terms of
the full gauge coupling for the specific brane as\footnote{See,
for example, \cite{polchinski} Vol. II, eq. (13.3.25).}:
\begin{equation} [g_{YM}^0]^2 \equiv e^{-\Phi}g_{YM}^2= 2\pi
\prod_{i=1}^3 2\pi\sqrt{\alpha'}L_i^{-1}\label{ym0}
\end{equation}
Thus the unitarity condition (\ref{unit}) appears with an extra
factor $[g_{YM}^0]^2$ on the right-hand side  (RHS) of the
equation.

When evaluating the four-point amplitude we chose to picture
change the vertex operators for both  $\chi$ fields, which in turn
ensures that there is no contribution from the internal part of
the  fermionic stress energy contribution. The upshot is the
following form  of the amplitude:
\begin{eqnarray} \label{four}
S_4(k_1,k_2,k_3,k_4)=
i{{C_{D2}g_0^4}\over{2\alpha'}}(2\pi)^4\delta^{(4)}(\sum_{i=1}^4k_i)
4\alpha'^2k_1\cdot k_3 \phantom{more space to the right}\\
\nonumber \left(\int_0^1\, dx x^{-\alpha' s-1}(1-x)^{-\alpha' t-1}
\prod_{_j=1}^3[x(1-x)]^{\nu_j(1-\nu_j)}Z_j(x) + s
\longleftrightarrow t\right)\,
\end{eqnarray}
where the $Z_j$ is the four-twist amplitude defined  in (\ref{Zo})
with $\nu=\nu_j$ while $s,t$ are the Mandelstam variables.

In order to compare the LHS and RHS of  (\ref{unit})  and thus
determine $const.$ we shall evaluate the amplitude  (\ref{Zo}) in
the limit $x\to 0$ first. As $x\to 0$, $F(x)\sim B(\n,1-\n)$ and
$F(1-x)\sim
    -\log(x/\d)$, where $\log\d\equiv
    2\psi(1)-\psi(\n)-\psi(1-\n)$ and $\psi(z)\equiv
    d\log\G(z)/dz$. Therefore, to take the limit $x\to 0$ we must
    do a Poisson resummation over $r_2$.
    This gives
    \beqa
    Z(x)=\frac{\p\sqrt{2\a'}}{L_2}\frac{{\rm const.}}{\sqrt{\sin(\p\n)}}[x(1-x)]^{-\n(1-\n)}F(x)^{-1}\\ \NO
    \sum_{m_1,m_2}\exp-\p\frac{F(1-x)}{F(x)}\left[\left(\frac
    {m_1L_1}{\pi\sqrt{2\alpha'}}\right)^2\sin(\p\n)+\left(\frac{\p m_2\sqrt{2\alpha'}}{L_2}\right)^2
    \frac{1}{\sin(\p\n)}\right]\sim \\ \NO
    \frac{2\p\sqrt{\a'}}{L_2}\frac{{\rm const.}\sqrt{\sin(\p\n)}}{\sqrt{2}\p}x^{-\n(1-\n)}
    \sum_{m_1,m_2}\left(\frac{x}{\d}\right)^{\left[
    \left(\frac{m_1L_1\sin(\p\n)}{\p\sqrt{2\a'}}\right)^2+\left(\frac{\p m_2\sqrt{2\a'}}{L_2}\right)^2\right]}\eeqa

The pre-factors $2\p\sqrt{\a'}/L_2$ from each of
 the the $Z_j$ contribution combine precisely into
 $[g_{YM}^0]^2/2\p$ for brane 2 (see eq.(\ref{ym0})).
 Therefore the  contribution of $g_{YM}^{0}$ on both sides of eq.
 (\ref{unit}) cancels.
Evaluating the four-point amplitude near $x=0$ yields a pole
associated with $s$-channel exchange of the corresponding gauge
field. Equating the LHS and RHS  of (\ref{unit}) in turn then
determines:\begin{equation}
const.=2\pi\prod_{j=1}^3\frac{\sqrt{2}\p}{\sqrt{\sin(\p\nu_j)}}\, .
\label{con}
\end{equation}

The  limit $x\to 1$ gives a  contribution from the $t$ channel
exchange of gauge bosons   associated with  brane 1. In this case
the resummation is over $r_1$  in (\ref{Zo}) which again
 factorizes to $[g_{YM}^0]^2$ associated with brane 1 in the four- point amplitude
  (\ref{four}) and  thus cancels the same gauge coupling contribution
  on the RHS of (\ref{unit}). Of course the rest of the calculation is consistent with
 the values of $const.$  in  (\ref{con}).

 \subsection{Generalization of the Lattice Summation}

 In the amplitude (\ref{Zo}) we assumed that the four twist
 fields were coming from the {\em same} intersection and therefore the
 summation over all possible parallelograms reduced simply to a
 sum over multiples of the lengths, $L_1$ and
 $L_2$, of the two cycles the branes wrap. We would like to generalize this amplitude
 to four twist fields coming from more than just one intersection.

 First let us consider the correlation function of a
 twist-antitwist pair from intersection i and a twist-antitwist
 pair from intersection j. Obviously, the fields coming from the
 same intersection must be separated by a lattice translation and
 therefore the distance between them is again a multiple of the
 length of one of the two cycles, namely $L_1$ or $L_2$. However,
 the minimum distance between fields from different intersections
 is not zero. In particular, it depends on the total number $I_{12}$ of
 intersections between the two branes and the lengths of the
 one-cycles they wrap as we show next.

 By translating the one-cycles by all possible lattice vectors in
 the covering space $\mathbb{C}$ of $T^2$ one observes that along
 one complete cycle each fixed point appears only once and that
 the one-cycle is divided into $I_{12}$ equal intervals of length
 $L/I_{12}$, where  $L$ stands for either $L_1$ or $L_2$ depending on the cycle under consideration.
 Therefore, the minimum distance between two different
 intersections is generally an integer multiple of $L_1/I_{12}$ or
 $L_2/I_{12}$. One must first decide on the labelling of the
 $I_{12}$ intersections. There are obviously two evident but
 equivalent labellings, namely, we can index the
 intersections in increasing order, starting from 0, along cycle 1 or cycle 2. Let us, for concreteness,
 label them along cycle 1. The
 minimum distance between intersections i and j on cycle 1 is then obvious, but the minimum
 distance between the same intersections on cycle 2 is not
 because the intersection points are ordered differently along
 this cycle. So one must first determine how the fixed points are
 ordered along the second cycle.

\begin{figure}
\begin{center}
\scalebox{0.7}{\rotatebox{-90}{\includegraphics{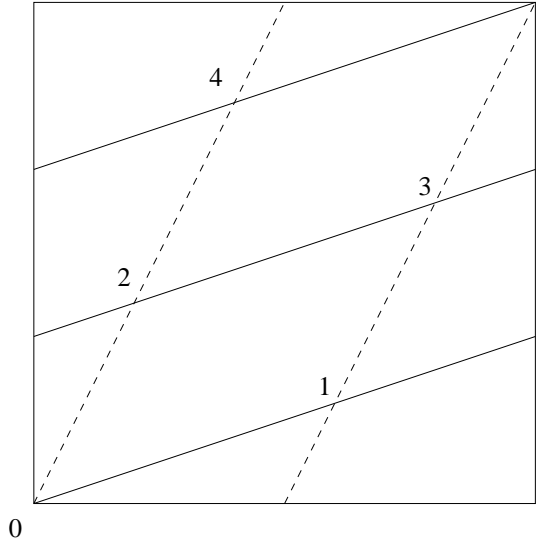}}}
\end{center}
\caption[]{\small The fundamental domain for a brane with wrapping
numbers (3,1) (solid line) and a brane with wrapping numbers (1,2)
(broken line). There are five intersection points labelled in
increasing order starting from 0 along the solid brane. Starting
from 0 and moving along the second brane (broken line) one meets
first fixed point 2. This is the integer k that generates the
automorphism (\ref{auto}) in this example. } \label{intersections}
\end{figure}

 To this end consider the cyclic group of order $I_{12}$, namely
 \beq
 \mathbb{Z}_{I_{12}}\equiv\{e,c,c^2,\ldots,c^{I_{12}-1}|c^{I_{12}}=e\}\eeq
 where $c$ is the generator of the group. To each fixed point we
 can uniquely associate an element of this group by the rule \beq
 j\longleftrightarrow c^j\eeq It can then be shown that the
 ordering of the fixed points along cycle 2 is given by the
 automorphism \beq
  g\longmapsto g^k,\,\,\forall g\in \mathbb{Z}_{I_{12}}\label{auto}\eeq where k is an integer between 1 and $I_{12}-1$
 which depends on the wrapping numbers of the two cycles. For this
 map to be an automorphism obviously k must not divide $I_{12}$
 for then the map is not injective. A general expression for k as
 a function of the four wrapping numbers has proved difficult to
 find so far apart for special cases of wrapping numbers. We
 emphasize that whatever this expression might be it must
 guarantee that k does not divide $I_{12}=n_1m_2-n_2m_1$. Until
 such an expression is known one can always determine this number
 k by drawing the cycles in the fundamental domain of the torus. k
 is given by the fixed point closest to 0 along cycle 2 (see
 Fig.\ref{intersections}).

\begin{figure}
\begin{center}
\scalebox{0.7}{\rotatebox{-90}{\includegraphics{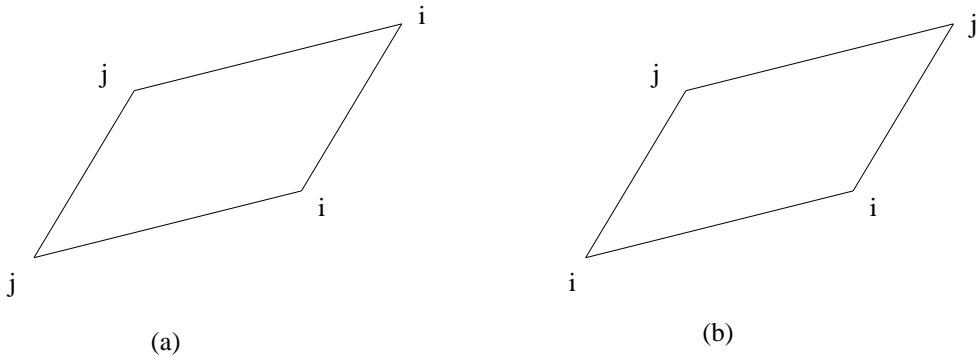}}}
\end{center}
\caption[]{\small The two configurations for a twist-antitwist
pair at intersection i and a twist-antitwist pair at intersection
j. Both configurations must be included in the string amplitude. }
\label{latticediagrams}
\end{figure}

The four-point function with twist fields from intersections i and
j gets contributions from the two lattice configurations in Figure
\ref{latticediagrams}. If $d_1(i,j)\propto L_1/I_{12}$ and
$d_2(i,j)\propto L_2/I_{12}$ are the minimal distances between
fixed points i and j along cycles 1 and 2 respectively, the
four-point amplitude takes the form
\begin{eqnarray}\label{four2}
S_4(k_1,k_2,k_3,k_4)=
i{{C_{D2}g_0^4}\over{2\alpha'}}(2\pi)^4\delta^{(4)}(\sum_{i=1}^4k_i)
4\alpha'^2k_1\cdot k_3 \phantom{more space right}\\
\nonumber \left(\int_0^1\, dx x^{-\alpha' s-1}(1-x)^{-\alpha' t-1}
\prod_{_j=1}^3[x(1-x)]^{\nu_j(1-\nu_j)}Z_j^{(1)}(x)+\right.\\\NO
\left. \int_0^1\, dx x^{-\alpha' t-1}(1-x)^{-\alpha' s-1}
\prod_{_j=1}^3[x(1-x)]^{\nu_j(1-\nu_j)}Z_j^{(2)}(x)\right),
\end{eqnarray}
 where
 \beqa
     Z^{(1)}(x)=const.[x(1-x)]^{-\n(1-\n)}[F(x)F(1-x)]^{-1/2}\phantom{more space to the right}\\
     \NO
     \sum_{r_1,r_2}\exp-\frac{1}{2\p\a'}\sin(\p\n)F(x)F(1-x)\left[\left(\frac{r_1L_1}{F(x)}\right)^2+
    \left(\frac{d_2(i,j)+r_2L_2}{F(1-x)}\right)^2\right]\eeqa and
 \beqa
     Z^{(2)}(x)=const.[x(1-x)]^{-\n(1-\n)}[F(x)F(1-x)]^{-1/2}\phantom{more space to the right}\\
     \NO
    \sum_{r_1,r_2}\exp-\frac{1}{2\p\a'}\sin(\p\n)F(x)F(1-x)\left[\left(\frac{d_1(i,j)+r_1L_1}{F(x)}\right)^2+
    \left(\frac{r_2L_2}{F(1-x)}\right)^2\right].\eeqa

The summation of these two lattice contributions in the case of
i=j gives the two terms in the amplitude (\ref{four2}). For
distinct i and j, however, there is no t-channel massless exchange
since twist fields from different intersections do not couple.
This can be seen explicitly from the string amplitude. As $x\to 0$
or $x\to 1$ only one of the terms  gives a massless exchange after
Poisson resummation. In particular, for the  $x\to 0$ limit one
needs to do a Poisson resummation in $r_2$ to see that only the
$Z^{(1)}$ term survives in this limit. The second term, which contains
$d_1(i,j)$, goes to zero even for $r_1=0$ in this limit.
Analogously, only the $Z^{(2)}$ term contributes in the $x\to 1$
limit.

To determine the overall normalization of the amplitude we proceed
as in the previous subsection. Instead of two s-channel poles and
two t-channel poles from gauge bosons living respectively on brane
1 and 2, we now get in the amplitude (\ref{four}) just two
s-channel poles, one for each type of gauge bosons. The
normalization constant is still given by equation (\ref{con}),
however.

One can ask if similar results hold for four-point amplitudes of
twist fields coming from more than two intersections. Clearly an
amplitude of a twist-antitwist pair from intersection i with a
twist from intersection j and an antitwist from intersection k is
not possible since the fields coming from intersection i must be
separated by a lattice translation which forces k=j. However,
four-point amplitudes of fields coming from four different
intersections are possible. In this case there will be a minimum
non-zero distance between each pair of twist fields which will
depend on the particular brane configuration. At most one lattice
configuration exists for a given set of twist fields all coming
from different intersections. The necessary and sufficient
condition for a non-vanishing amplitude of two twist fields from
intersections i and j and two antitwist fields from intersections
k and l is i-k=l-j. Such amplitudes do not contain massless
exchanges though and so their overall normalization cannot be
determined directly by the above method. Nevertheless, this
normalization constant is part of the quantum amplitude, which is
independent of the global effects of the lattice, and hence it
must be also given by (\ref{con}).

\section{Four and three-point functions with two independent angles}

The above method can be directly applied to the problem of a
four-point amplitude with two independent angles
(Fig.\ref{nonparallel}).  The boundary conditions now read
\beqa\label{bc1}
\partial X+\bar{\partial}\bar{X}=0,\,\,
\partial\bar{X}+\bar{\partial}X=0,\,\,\,{\rm on}\,\,\,
(-\infty,x_1)\cup(x_2,x_3)\cup(x_4,+\infty) \\ \NO
e^{i\pi\nu}\partial X+e^{-i\pi\nu}\bar{\partial}\bar{X}=0,\,\,
e^{-i\pi\nu}\partial\bar{X}+e^{i\pi\nu}\bar{\partial}X=0,\,\,\,{\rm
on}\,\,\, (x_1,x_2)\\ \NO e^{-i\pi\l}\partial
X+e^{i\pi\l}\bar{\partial}\bar{X}=0,\,\,
e^{i\pi\l}\partial\bar{X}+e^{-i\pi\l}\bar{\partial}X=0,\,\,\,{\rm
on}\,\,\, (x_3,x_4)
 \eeqa In the appendix we evaluate the quantum amplitude
 $\langle\s_\n(x_1)\s_{-\n}(x_2)\s_{-\l}(x_3)\s_{\l}(x_4)\rangle$.
 The result is \beq Z_{qu}(x)={\rm
    const.}x^{-\n(1-\n)}(1-x)^{-\n\l}I(x)^{-1/2}\eeq where \beq
    I(x)\equiv (1-x)^{1-\n-\l}\left[B(\n,\l)F_1(1-x)K_2(x)+B(1-\n,1-\l)F_2(1-x)K_1(x)\right
    ].\label{Ix}.\eeq $B(\n,\l)$ is the Euler Beta function and
    $F_i$, $K_i$ are Hypergeometric functions defined in the
    appendix.

     From the boundary conditions (\ref{bc1}) we determine the classical
    solutions  \beqa
    \partial
    X(z)=a\om_{1-\n,\l}(z)\equiv\tilde{a}e^{i\p(\l-1)}\tilde{\om}_{1-\n,\l}(z)\\\NO
     \bar{\partial}
    \bar{X}(\bar{z})=-a\om_{1-\n,\l}(\bar{z})\equiv-\tilde{a}e^{-i\p(\l-1)}\tilde{\om}_{1-\n,\l}(\bar{z})\\\NO
     \partial
    \bar{X}(z)=b\om_{\n,1-\l}(z)\equiv\tilde{b}e^{-i\p\l}\tilde{\om}_{\n,1-\l}(z)\\\NO
     \bar{\partial}
    X(\bar{z})=-b\om_{\n,1-\l}(\bar{z})\equiv-\tilde{b}e^{i\p\l}\tilde{\om}_{\n,1-\l}(\bar{z})\eeqa

    Again the parameters $a$ and $b$ are arbitrary and reflect the freedom in specifying the lengths $d_1$ and $d_2$ of the
    four-sided polygon. However, to obtain a three-point amplitude one must take the limit  $x_2\rightarrow x_3$. Unless
    $1-\n-\l=0$, which is precisely the case of one independent
    angle considered above, one of the two linearly independent solutions becomes singular in this limit. For $1-\n-\l>0$, as we will assume
    without loss of generality,
    $\om_{\n,1-\l}(z)=(z-x_1)^{-\nu}(z-x_2)^{\nu-1}(z-x_3)^{\l-1}(z-x_4)^{-\l}$
    develops a non-integrable singularity at $z=x_3$ in the limit
    $x_2\rightarrow x_3$.  Therefore, the four-point amplitude that reduces to the three point amplitude
    must have $b=0$. This is to be expected since the distance $d_2$ cannot be an independent parameter if one wants
    to get a three-point amplitude. In fact if  $b$ is set to zero $d_2$ becomes a function of $d_1$ and $x_2$ which tends to zero as $x_2\to x_3$ as required.
    To keep the discussion general though we first consider the problem with arbitrary $a$ and $b$.

    The classical action is given by \beqa
    S_{cl}=-\frac{1}{4\p\a'}\left[\tilde{a}^2\int_{\mathbb{C}_+}d^2z|\tilde{\om}_{1-\n,\l}(z)|^2+\tilde{b}^2\int_{\mathbb{C}_+}d^2z|\tilde{\om}_{\n,1-\l}(z)|^2\right]
    =\\ \NO -\frac{I(x)}{4\a'}\left[\tilde{a}^2+(1-x)^{-2(1-\n-\l)}\tilde{b}^2\right]\eeqa where the integrals have been
    evaluated using the method of \cite{KLT} to
    factorize closed string amplitudes into product of open string
    amplitudes. To determine the coefficients $\tilde{a}$ and $\tilde{b}$
    we impose the global monodromy conditions
    \beq\int_{C'_1}dX^2=d_1,\,\,\,\,\int_{C'_2}dX^2=d_2\eeq where the contours $C'_1$ and
    $C'_2$ are shown in Figure \ref{contours}b.
     The two other lengths of the polygon are
    automatically determined and no other contours provide any additional
    information\footnote{Note, however, that these two conditions become linearly dependent when $1-\n-\l=0$ because $d_1=d_2$. In that case, which is the problem of one independent angle discussed earlier, one should take the contours of Fig.\ref{contours}a.}.
    Solving these conditions for $\tilde{a}$ and
    $\tilde{b}$ we obtain \beq \tilde{a}=i\left[
    (1-x)^{-(1-\n-\l)}B(\n,\l)F_1(1-x)d_1+B(1-\n,1-\l)F_2(1-x)d_2\right]/J(x)\eeq
    and
    \beq \tilde{b}=i\left[
    B(\n,\l)F_1(1-x)d_2+(1-x)^{(1-\n-\l)}B(1-\n,1-\l)F_2(1-x)d_1\right]/J(x)\eeq
    where \beq J(x)\equiv
    (1-x)^{-(1-\n-\l)}\left[B(\n,\l)F_1(1-x)\right]^2+(1-x)^{(1-\n-\l)}\left[B(1-\n,1-\l)F_2(1-x)\right]^2.\eeq

    Although the general four-point amplitude with arbitrary $d_1$
    and $d_2$ is interesting itself, we want to evaluate the
    three-point amplitude which gives the instanton corrections to the Yukawa
    couplings. $b$ then must be set to zero and the monodromy
    conditions give instead \beq
    \tilde{a}=\frac{id_1}{B(\n,\l)F_1(1-x)}\eeq while $d_2$
     is
    $x$-dependent \beq
    d_2(x)=d_1(1-x)^{1-\n-\l}\frac{B(1-\l,1-\n)F_2(1-x)}{B(\l,\n)F_1(1-x)}\rightarrow
    0\eeq as $x\rightarrow 1$ which therefore correctly produces a
   three-point amplitude in this limit. The on-shell action becomes \beq
    S_{cl}=\frac{I(x)}{4\a'}\left(\frac{d_1}{B(\n,\l)F_1(1-x)}\right)^2\eeq
    and the complete four-point amplitude takes the form \beq
    Z_4(x)={\rm const.}x^{-\n(1-\n)}(1-x)^{-\n\l}I(x)^{-1/2}\sum_m\exp-\frac{I(x)}{4\a'}\left(\frac{d_1(m)}{B(\n,\l)F_1(1-x)}\right)^2.\eeq
    Here $d_1(m)=L_0+m L$ is the most general form the distance
    $d_1$ can take when the polygon is embedded in a lattice.
    $L_0$ and the cycle length $L$ will generically depend on the lattice and the
    D-brane configuration. The overall normalization constant can be determined by the same method
    used to fix the normalization of the four-point amplitude with one independent
    angle. We specialize to $T^6$ since the amplitude of most interest is the one
    involving factorizable three-cycles on $T^6$. We find \beq
    {\rm const.}=16\p^{5/2}.\eeq

    The limit $x\to 1$ of the four-point
    amplitude produces the full expression for the three-point
    amplitude. In particular, since the conformal weights of the twist fields satisfy $h_\n+h_\l-h_{\n+\l}=\n\l$, the latter
    contains the correct singularity in this limit in agreement
    with the operator product expansion \beq
    \mathcal{O}_i(z_1)\mathcal{O}_j(z_2)\sim\sum_kC_{ij}^k\mathcal{O}_k(z_2)(z_2-z_1)^{h_k-h_j-h_i}.\eeq
    One can then show that the full three-point amplitude for branes wrapping factorizable three-cycles on $T^6$ takes
    the form\footnote{See Note Added at the end of the paper.}
    \beqa\label{final}Z_3=2\p\prod_{j=1}^3\left[\frac{16\p^2 B(\n_j,1-\n_j)}{
    B(\n_j,\l_j)B(\n_j,1-\n_j-\l_j)}\right]^\frac14\sum_m\exp-\frac{A_j(m)}{2\p\a'}=\\ 
    2\p\prod_{j=1}^3\left[\frac{16\p^2\G(1-\n_j)\G(1-\l_j)\G(\n_j+\l_j)}{ \G(\n_j)\G(\l_j)\G(1-\n_j-\l_j)}\right]^\frac14\sum_m\exp-\frac{A_j(m)}{2\p\a'}\NO
    \eeqa
    where $A_j(m)$ is the area of the triangle formed by the three
    intersecting branes on the j-th torus. Note that the amplitude is completely
    symmetric in all three angles of the triangle as it should.

 The above  three-point correlation function of bosonic twist fields
 (\ref{final}) is the key contribution to the  physical Yukawa coupling
of two fermionic and one bosonic field. The full three-point
amplitude of these fields  can  be determined by employing the
normalization factor for the disk amplitude,
$C_{D2}=1/(g_0^2\alpha')$, and the bosonic string vertex operator
in -1 picture [eq.(\ref{bos})], $g_0$, as determined in Section
2.4. In addition, the normalization factor of the fermionic vertex
operators in -1/2 picture [eq.(\ref{fer})] turns out to be
$2^{\textstyle{1\over 4}}\, \sqrt{\alpha'}\,g_0$. [This
normalization can be determined from the string amplitudes of
fermionic fields to gauge vector bosons, along the same lines as
described for the corresponding bosonic fields in Section  2.4.]
Thus the final expression for the physical Yukawa couplings is
given by $\sqrt{2}\,g_0 \times  Z_3$.

    A comprehensive analysis of the triangles that contribute to
    this amplitude for a given set of intersections in diverse brane configurations and
    including the effect of non-trivial complex structure or Wilson lines has been
    done in \cite{CIM}.

    \section{conclusions}

    We have applied conformal field theory  techniques to obtain three-point and
    four-point correlation functions of twist fields from
    D-branes wrapping  factorizable n-cycles of $T^{2n}=T^2\times T^2\cdots$ and intersecting at points
    in  the internal $T^{2n}$.  The method allows for a complete
    determination of the amplitude including the quantum
    contribution.  Its most interesting  application  is to
     the three-point function  calculation of intersecting D6-branes wrapping factorizable
     three-cycles of $T^6$, which in turn gives
    the complete  Yukawa coupling of  two four-dimensional chiral
    fermions  to the Higgs field.

  The method also applies to the study of the
     three-point and four-point
    twist field correlation functions in models with orientifold
    and orbifold projections, as discussed in Subsection II C.  Due to the mirror symmetry projection
    along an $n$-plane in $T^{2n}$ the branes can now
      intersect with an orientifold image of another brane.
      However the amplitude calculation for states at the
      intersection proceeds analogously.
      As for the $Z_2\times Z_2$ orbifold projection, we mention that the combined orbifold and orientifold action maps cycles into
themselves; thus each intersection is accompanied by  a
combination of the  orientifold and orbifold images that have to
be carefully taken into account. We hope to return to the detailed
discussion of these contributions in the future work.


    The formalism   can  be applied to the calculation of three and four
    fermion couplings of  classes of  Type II   orientifold   compactifications  with
    intersecting D-branes.  Among them the  four-dimensional Type IIA orientifold compactification
    with intersecting D6-branes is interesting; supersymmetric compactifications of this type are directly
    related to $G_2$ compactification of M-theory. In addition to the Yukawa couplings,
     the four-point amplitude of chiral fermions can
    provide low energy corrections to the effective four-fermion
    coupling
    and may be of phenomenological interest.  We plan to
    address further  these effects in concrete
    semi-realistic N=1 supersymmetric models such as those
    constructed in, e.g.,
    \cite{CSU1,CSU2,CPS,CP}.

\section*{Acknowledgments}

We would like to thank   Michael Douglas,  Paul Langacker, Gary
Shiu  and Angel Uranga for useful discussions. M.C. would like to
thank the New Center for Theoretical Physics at Rutgers University
and the Institute for Advanced Study, Princeton,
 for hospitality and
support  during the course of this work. Research supported in
part by DOE grant DOE-FG02-95ER40893, NATO linkage grant No. 97061
(M.C.) and National
 Science Foundation Grant  No. INT02-03585 and the  Fay. R. and Eugene L.
 Langberg Chair (M.C.).
\vspace{.5in}

{\it Note Added}

We are grateful to the authors of hep-th/0404134 for pointing out, after 
this article was published, that the correct
power of the normalization of the three point amplitude in (\ref{final}) is 1/4 and
not 1/2 as we had written in the original version.

\newpage
\appendix
\section{Evaluation of Quantum Four-Point Amplitude with Two
Independent Angles}

 Here we evaluate the four-point correlation function
 $\langle\s_\n(x_1)\s_{-\n}(x_2)\s_{-\l}(x_3)\s_{\l}(x_4)\rangle$.
 The calculation closely parallels the calculation for closed
 string amplitudes on orbifolds \cite{BKM}.
 As before we introduce the auxiliary correlators $g(z,w)$,
 $\bar{g}(z,w)$, $h(z,w)=-\bar{g}(z,w)$ and $\bar{h}(z,w)=-g(z,w)$
 as defined in section II.A. In terms of \beq
    \om_{\n,\l}(z)=(z-x_1)^{-\nu}(z-x_2)^{\nu-1}(z-x_3)^{-\l}(z-x_4)^{\l-1}\eeq
 these take the form \beq
 g(z,w)=\om_{1-\n,\l}(z)\om_{\n,1-\l}(w)\left(\frac{P(z,w)}{(z-w)^2}+A(\{x_i\})\right)\eeq\beq
 h(\bar{z},w)=\om_{\n,1-\l}(\bar{z})\om_{\n,1-\l}B(x_i) \eeq where
 \beq \label{cond}P(z,w)=\sum_{i,j=0}^2a_{ij}w^iz^j \eeq and the condition \beq
 g(z,w)\sim (z-w)^{-2}\eeq determines all coefficients $a_{ij}$
 except for $a_{20}$, $a_{11}$ and $a_{02}$ for which it provides
 two linear equations. Solving these for $a_{02}$ and $a_{11}$
  we find \beqa \langle T(z)\rangle=\lim_{w\rightarrow
    z}\left(g(z,w)-\frac{1}{(z-w)^2}\right)
    =\frac{A(\{x_i\})+z^2+a_{21}z+a_{20}}{(z-x_1)(z-x_2)(z-x_3)(z-x_4)}\\
    \NO +\frac 12
    \n(1-\n)\left(\frac{1}{(z-x_1)}-\frac{1}{(z-x_2)}\right)^2+ \frac 12
    \l(1-\l)\left(\frac{1}{(z-x_3)}-\frac{1}{(z-x_4)}\right)^2\\
    \NO -\left(\frac{\n}{(z-x_1)}+\frac{1-\n}{(z-x_2)}\right)
    \left(\frac{1-\l}{(z-x_3)}+\frac{\l}{(z-x_4)}\right)\eeqa
    where $a_{21}=-[(1-\n)x_1+\n x_2+\l x_3+(1-\l)x_4]$. The
    freedom in $a_{20}$ does not affect the final result since
    $A(\{x_i\})$ will be determined after $a_{20}$ is fixed.
    Fixing  \beq
    a_{20}=\frac12[(1-\n+\l)x_1x_3+(1-\n-\l)x_1x_4-(1-\n-\l)x_2x_3+(1+\n-\l)x_2x_4]\eeq
    and using $SL(2,\mathbb{R})$ invariance to set $x_1=0,\,\,
    x_2=x,\,\, x_3=1,\,\, x_4\rightarrow\infty$ we find \beq
    \partial_x Z_{qu}(x)=\frac
    12\frac{(1-\n-\l)}{x-1}-\frac{\n(1-\n)}{x}-\frac{(1-\n)(1-\l)}{x-1}+\frac{A(x)}{x(x-1)}\eeq
    where \beq
    A(x)=\lim_{x_4\rightarrow\infty}-x_4^{-1}A(\{x_i\})\eeq
    and
    \beq
    Z_{qu}(x)=\lim_{x_4\rightarrow\infty}|x_4|^{\l(1-\l)}\langle\s_\n(0)\s_{-\n}(x)\s_{-\l}(1)\s_{\l}(x_4)\rangle\eeq

    As before $A(x)$ is determined by imposing the quantum
    monodromy conditions \beq
    \int_{C_i}dX=\int_{C_i}d\bar{X}=0\eeq leading again to
    conditions (\ref{qmonodromy1}) and (\ref{qmonodromy2}).
    We  insert \beq g(z,w)\rightarrow [\frac
    12(1-\n-\l)x+(1-\l)(z-x)+A(x)]\tilde{\om}_{1-\n,\l}(z)\eeq
    \beq \bar{g}(z,w)\rightarrow B(x)\tilde{\om}_{\n,1-\l}(z)\eeq
    and arrive at the two equations
    \beqa[\frac12(1-\n-\l)x+A(x)]\int_{C_1}dz\tilde{\om}_{1-\n,\l}(z)+(1-\l)\int_{C_1}dz(z-x)\tilde{\om}_{1-\n,\l}(z)\\ \NO -e^{-2\p
    i\n}B(x)\int_{C_1}dz\tilde{\om}_{1-\n,\l}(z)=0 \eeqa and
     \beqa[\frac12(1-\n-\l)x+A(x)]\int_{C_2}dz\tilde{\om}_{1-\n,\l}(z)+(1-\l)\int_{C_2}dz(z-x)\tilde{\om}_{1-\n,\l}(z)\\ \NO -
    B(x)\int_{C_2}dz\tilde{\om}_{1-\n,\l}(z)=0. \eeqa
    To solve these we need the integrals \beq
    \int_{C_1}dz\tilde{\om}_{1-\n,\l}(z)=e^{i\p(1-\n-\l)}B(\n,1-\n)K_1(x)\eeq
    \beq
    \int_{C_2}dz\tilde{\om}_{1-\n,\l}(z)=e^{i\p(1-\l)}B(1-\l,1-\n)(1-x)^{1-\n-\l}F_2(x)\eeq
    where \beq K_1(x)\equiv F(\n,\l;1,x)\eeq
    \beq K_2(x)\equiv F(1-\n,1-\l;1;x)\eeq
    \beq F_1(x)\equiv F(\n,\l;\n+\l,x)\eeq and \beq F_2(x)\equiv
    F(1-\n,1-\l;2-\n-\l,x).\eeq One also needs the following
    identities \beq
    (1-\n)(1-\l)F(\n,\l;2,x)=(1-\n-\l)K_1(x)+(1-x)\partial_x
    K_1(x)\eeq
    \beq
    (1-\n)(1-\l)F(1-\l,1-\n;3-\n-\l;1-x)=-(2-\n-\l)\partial_xF_2(1-x)\eeq
    \beq
    B(1-\n,1-\l)(1-x)^{1-\n-\l}F_2(1-x)=B(\n,\l)F_1(1-x)-B(1-\n,1-\l)B(\n,\l)(1-\n-\l)K_1(x)\eeq
    and \beq K_1(x)=(1-x)^{1-\n-\l}K_2(x).\eeq After some
    algebra we arrive at the desired result \beq
    2A(x)=x(1-x)\partial_x\log\left[B(\n,\l)F_1(1-x)K_2(x)+B(1-\n,1-\l)F_2(1-x)K_1(x)\right].\eeq
    Hence \beq Z_{qu}(x)={\rm
    const.}x^{-\n(1-\n)}(1-x)^{-\n\l}I(x)^{-1/2}\eeq where
    \beq
    I(x)\equiv (1-x)^{1-\n-\l}\left[B(\n,\l)F_1(1-x)K_2(x)+B(1-\n,1-\l)F_2(1-x)K_1(x)\right
    ].\label{Ix}\eeq For completeness let us also give the result for
    $B(x)$, namely \beq B(x)=\frac12 e^{-2\p
    i\l}x(1-x)^{(2-\n-\l)}\partial_x\log\left[\frac{B(\n,\l)F_1(1-x)K_2(x)+B(1-\n,1-\l)F_2(1-x)K_1(x)}{(K_2(x))^2}\right].\eeq

\end{document}